# Production of Metal-free Diamond Nanoparticles

Laia Ginés[1]*, Soumen Mandal[1], David John Morgan[2], Ryan Lewis[3], Phil Davies[2], Paola Borri[3], Gavin Morley[4] and Oliver A. Williams[1]*

[1]School of Physics and Astronomy, Cardiff University, Cardiff CF24 3AA, UK
[2]Cardiff Catalysis Institute, School of Chemistry, Cardiff University, Cardiff CF10 3AT, UK
[3]School of Biosciences, Cardiff University, Cardiff CF10 3AX, UK
[4] Department of Physics, University of Warwick, Coventry CV4 7AL, UK

KEYWORDS Nanoparticles, Nanodiamonds, Milling, Colloids, X-Ray Photoelectron Spectroscopy.

**ABSTRACT:** In this paper, the controlled production of high quality metal-free diamond nanoparticles is demonstrated. Milling with tempered steel is shown to leave behind iron oxide contamination which is difficult to remove. Milling with SiN alleviates this issue but generates more non diamond carbon. Thus the choice of milling materials is critically determined by the acceptable contaminants in the ultimate application. The removal of metal impurities, present in all commercially available nanoparticles, will open new possibilities towards the production of customized diamond nanoparticles, covering the most demanding quantum applications.

INTRODUCTION

Diamond nanoparticles have been highlighted as an important material for a wide range of applications. Due to bio-compatibility,[1,2] and non-cytotoxicity[3] of diamond, diamond nanoparticles have been used in biomedical applications such as optical bio-imaging[4] or drug delivery.[5,6] Diamond nanoparticles are also used as nucleation centres for synthetic diamond growth,[7] with potential application for nano electro mechanical systems (NEMS)[8,9] or quantum devices such as superconducting quantum interference devices (SQUIDs).[10] Another important application is the use of diamond nanoparticles as single photon sources.[11] Defects present in the diamond lattice or defects intentionally created (known as colour centres), can act as single photon emitters.[12] For instance, due to its room-temperature high photostability and narrow emission in the zero phonon line, the $SiV^-$ is a good candidate for quantum computing and quantum cryptography applications[13]. Another well-known colour centre, the $NV^-$, is interesting due to its spin properties.[14,15,16,17,18,19] Spin manipulation possibilities make this colour centre attractive as a magnetic field sensor and for magnetic imaging.[20] Furthermore, recent studies in levitated diamond nanoparticles containing the $NV^-$ centre have been proposed to detect quantum superposition states[21,22] and as a method to detect quantum gravity.[23,24]

Although a variety of commercial nanoparticles are available, increasingly demanding applications require the use of high quality contaminant-free diamond nanoparticles. Non-diamond carbon ($sp^2$) is commonly present in commercial diamond nanoparticles. The presence of such carbon on diamond surfaces leads to particle aggregation[25] and reactive $sp^2$ species are detrimental for biological applications. Additionally, commercial nanodiamonds typically have a high concentration of nitrogen defects [N]>100 ppm which reduces $NV^-$ spin coherence times and hence magnetic field sensitivity.[26] This high nitrogen content is also a problem for the levitation of diamond nanoparticles in high vacuum as the nitrogen absorbs the trapping laser light, heating and burning the diamond nanoparticles.[27,28] Furthermore, a non negligible amount of metal contaminants[29,30] is also present in commercially available nanoparticles, reducing the possibilities of the magnetic-related applications. Inductively coupled plasma mass spectrometry (ICP-MS) studies performed by *Volkov et al.*[29] over twenty different commercial detonation diamond particles, detected high amounts of metal impurities in all of them.

For these reasons, it is highly important to be able to create customized diamond particles. Several methods for particles' production are known up to date. The most versatile one is the production of diamond nanoparticles by crushing CVD diamond or bulk diamond using milling techniques.[31,32] This approach enables particles size distribution control and offers the possibility of creating custom colour centres as well as particles from bulk diamond with low nitrogen concentration. Nevertheless, deep cleaning methods are required after the milling process.

In this paper, the production of metal free particles from commercial single crystal bulk diamond is shown. Different milling materials are compared.

EXPERIMENTAL METHODS

Commercial CVD grown single crystal (SC) diamond samples 0.3 mm thick (2.6 mm × 2.6 mm sized) sourced from Element Six, were used in this study. Two different grinding bowls, one made from tempered steel and one made from silicon nitride, were used for crushing the SC plates. 12 SC plates (95 mg approximately) were introduced in each grinding bowl with 5 ml of DI water and 30 and 40 grams of silicon nitride and tempered steel grinding balls (d=3mm) respectively. Samples were milled in the Planetary Micro Mill Pulverisette 7, following 6 cycles of 5 min on/15 min off at 1100 rpm (~95 g). After the milling process, the samples were cooled down and taken out of the grinding bowls and several acid cleaning processes were performed. For iron removal, the cleaning was performed as described by *Heyer et al.*[32] For the silicon nitride cleaning, 20 ml of sample were mixed with 30 ml of orthophosphoric acid ($H_3PO_4$) and the mixture was stirred continuously in a condenser during 24 hours at 180°C bath temperature. To remove the acids, both solutions underwent repeated washing and centrifugation cycles at 30000g, removing the supernatant after each centrifugation process and adding deionized water to the pellet until the pH reached a value between 5.8- 6.0. The slurries were dried in a hot plate to obtain the powders. For the silicon nitride grinding process, the obtained powder was introduced again in the condenser after being dissolved in 20 ml of water, and 30 ml of concentrated sodium hydroxide (NaOH) solution was added. The mixture was stirred in the condenser during 24 hours at 150°C. The washing and centrifugation cycles were repeated as previously described until pH 5.8-6 was reached. After the cleaning and centrifugation cycles, the tempered steel milled and the silicon nitride milled powders were treated in a furnace under air atmosphere at 600°C for 5 hours. Different aqueous colloids were prepared from the treated powders by dispersing 0.01 g of powder in 20 ml of deionized water. The colloids were dispersed via ultrasound and the solutions were centrifuged at different accelerative forces (5000 g, 10000 g, 20000 g and 30000 g) at 10°C in a Sigma 3-30 KS centrifuge. Dynamic light scattering (DLS) and nanoparticle tracking analysis (NTA) measurements were performed to measure the particles' size distribution. The Malvern Zetasizer Nano ZS used in our experiments was equipped with a 633nm laser in backscattering configuration (173°) and the Malvern Nanosight LM10 was equipped with a 635 nm laser.

Raman measurements were recorded in an *inVia* Renishaw confocal Raman microscope equipped with a 532 nm laser. All the measurements were acquired using the same parameters: 10 seconds acquisition time and 50 accumulations.

For comparison of the surface chemistry of each powder, X-ray photoelectron spectroscopy (XPS) measurements were performed in a Thermo Scientific K-Alpha+ spectrometer. Spectra were acquired using a monochromatic Al source operating at 72 W (6 mA emission current x 12 kV anode potential). A survey and high resolution spectrum were acquired at pass energies of 150 eV and 40 eV respectively. Charge neutralization was achieved using the K-Alpha charge neutralization system, employing a combination of both electrons and low energy argon ions. All XPS spectra were calibrated with the carbon C1s peak at 285 eV.

RESULTS AND DISCUSSION

XPS measurements were taken in various steps. 1) On pristine SC substrate, 2) after tempered steel milling, 3) after tempered steel milling followed by acid cleaning and 4) after the silicon nitride milling followed by its corresponding acid cleaning treatment (see experimental methods). Figure 1 shows the XPS survey for the mentioned processes with all the detectable elements present in the samples.

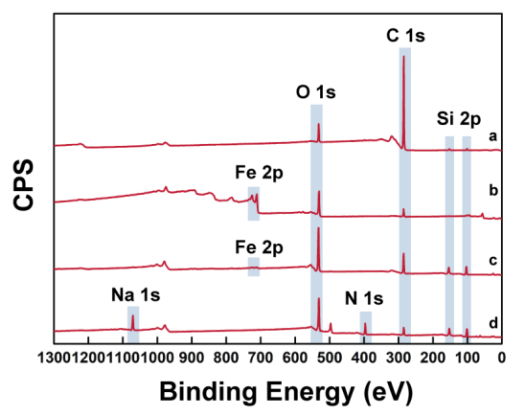

Figure 1. Survey XPS spectrum of the different samples. a) SC raw material just before the milling, b) powder after milling using tempered steel grinding bowl, c) powder after the tempered steel milling and the acid cleaning to remove the metal contaminants and d) powder after the silicon nitride milling.

Figure 1a shows the XPS spectrum of the as received SC substrate. The XPS spectrum shows two clear peaks corresponding to carbon (C1s peak) and oxygen (O1s peak) elements at 285 eV and 531.8 eV respectively. A small third peak (Si2p) also appears at lower binding energies. No silicon should be present in the sample, so the sample was subjected to argon cluster cleaning. The complete removal of the silicon after the cluster cleaning indicates some kind of surface contamination. (See Figure S1 Supporting Information). The survey XPS spectrum in figure 1b, corresponds to the sample milled with the tempered steel milling bowl and balls. The spectrum shows four elements: carbon (C1s), oxygen (O1s), iron (Fe2p) and silicon (Si2p). The Fe2p peak confirms the presence of iron in the sample produced in the milling process. Although acid cleaning was performed in order to remove all the metallic components, non-negligible amounts of iron were detected (XPS detection limit is 10 ppm) after the acid cleaning, as shown in figure 1c. The Si2p peak presence can be neglected as this silicon is due to contamination present previous to the milling process and can be removed with argon ion cleaning inside the XPS chamber.

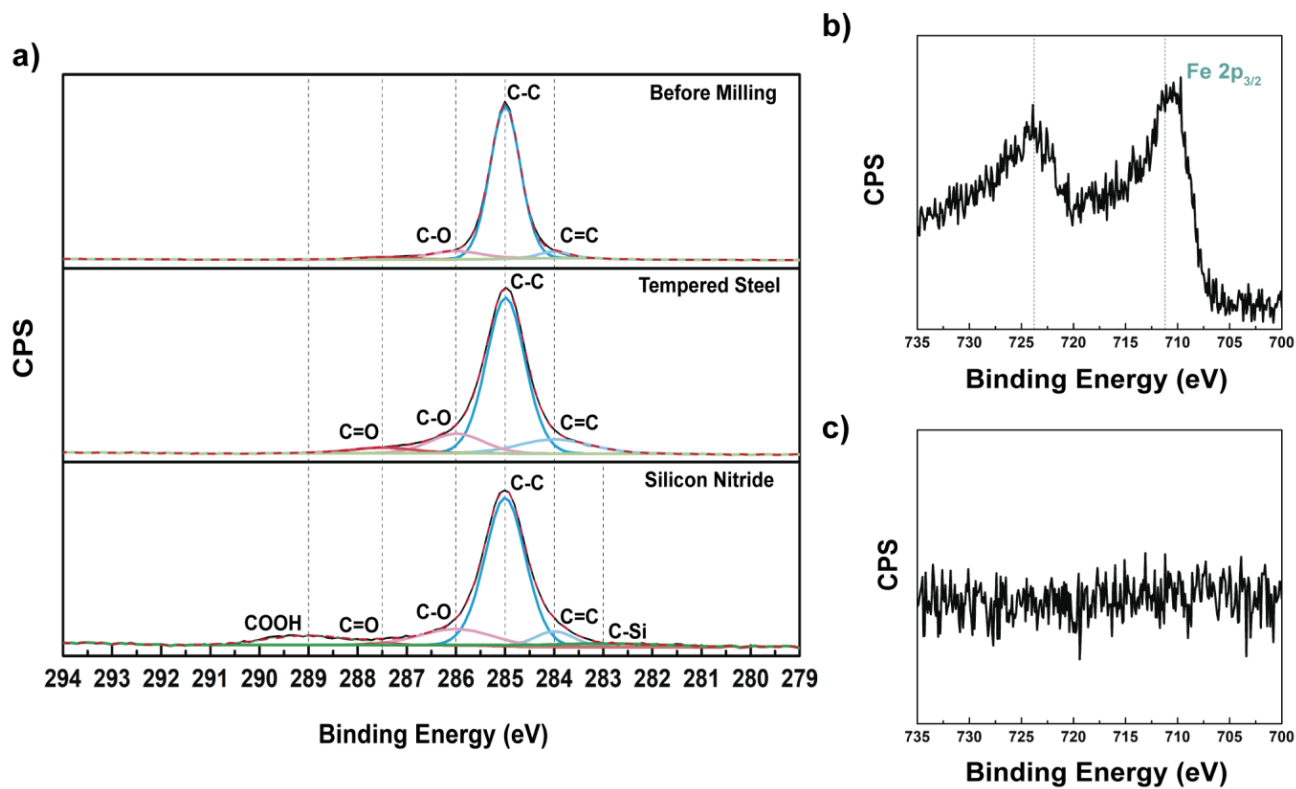

Figure 2. High resolution XPS scans for the carbon and iron elements. a) Carbon (C1s) deconvoluted peak for the as received SC, and the powders after been milled with the tempered steel and the silicon nitride grinding bowls respectively. b) Iron (Fe 2p) peak for the SC samples milled using a tempered steel grinding bowl and c) Fe 2p peak of the SC sample milled with the silicon nitride grinding bowl.

The C1s, O1s, Si2p, N1s and Na1s peaks are present in the SC sample milled with silicon nitride (shown in Figure 1d).

In this case, the Si2p and N1s peaks are related to silicon and nitrogen contamination due to insufficient cleaning of the silicon nitride produced in the milling. The Na1s peak is due to sodium contamination in the cleaning process, as NaOH was used in the cleaning. Nevertheless, iron (Fe2p peak) was not detected making the use of a silicon nitride grinding bowl extremely important in magnetometry applications.

The atomic percentages of the different elements present in the samples can be calculated from the survey spectrum, after the normalization of the peak areas considering the appropriate sensitivity factors (see table S1 in Supporting Information).[33] A drastic increase in the O1s/C1s ratio is observed after the milling processes because the samples are subjected to an air annealing (oxidation) treatment after the respective acids cleaning treatments. Further increase in the O1s/C1s ratio was obtained for the SC sample milled with the silicon nitride.

To obtain further information about the elements present in the samples, high resolution scans were performed in the region of interest. The data was fitted using Gaussian fits in CasaXPS, after subtraction of a Shirley type background.

All the peaks present in the XPS survey were analysed in detail (figures S2, S3 and S4 Supporting Information), but only the carbon and the iron spectra are discussed in this paper. In figure 2, high resolution scans for the C1s and the iron Fe2p peaks of the different samples are shown. The fitting of the C1s peak shows different components with small binding energy shifts between them. However, there is no consensus in the literature to assign each binding energy (BE) to a component, as different BE values have been reported for the same chemical species.[34,35,36,37] BE shifts of -1,+1,+2.5 and +4 eV were taken from the literature[38] and assigned to graphitic carbon, hydroxyl (C-OH)/ether (C-O-C), carbonyl (C=O) and carboxyl groups (COOH). An additional shift of -3 eV was considered for the silicon nitride milled sample, corresponding to Si-C bonds.[39,40]

The C1s peak for the SC sample before any milling process shows a symmetric spectrum with a dominant peak at 285 eV, attributed to the C-C bond (sp$^3$ bonded carbon). In the same graph, small contributions attributed to both C=C bonds and C=O bonds were also observed at lower and higher binding energies respectively. The SC diamond sample milled with tempered steel, presents however, a slightly asymmetric peak consisting of four Gaussian peaks, centered at 284 eV, 285 eV, 286 eV and 287.5 eV. The first peak can be assigned to sp$^2$ carbon (C=C). The peak at

285 eV corresponds to sp$^3$ bonded carbon, the peak at 286 eV is attributed to –C-H/C-O bonds and the peak at 287.5 eV is assigned to the C=O bonds.[41]

The SC sample after the silicon nitride milling presents in contrast, an asymmetric peak with a tail towards higher binding energies, which indicates a higher sp$^2$ carbon concentration. Apart from the peaks described for the tempered steel sample, two more peaks are clearly observed.

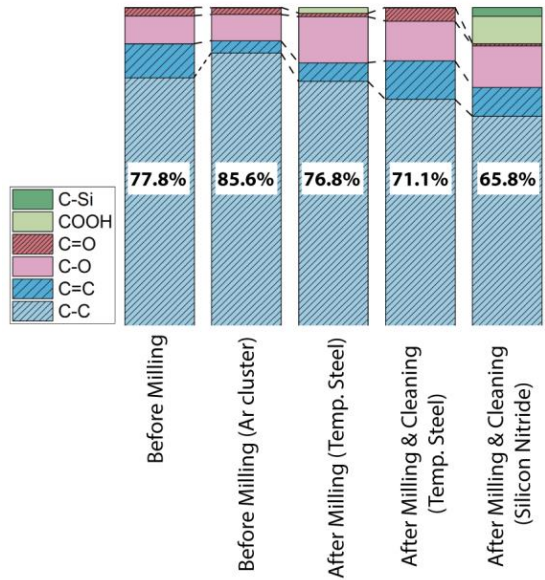

Figure 3. Bar plot with the percentages of the relative contents of the C1s peak for the different samples.

The peak at the lowest energy, 283 eV confirms the presence of carbon bonded to silicon, and the peak at 289 eV can be assigned to carboxyl groups (COOH) or to the π-π* transition.[42,43]

The differences between the three samples are more pronounced representing the percentage of the relative contents as shown in figure 3. The sample before the milling process subjected to the cluster argon ion cleaning has also been included for comparison.

A reduction in the sp$^3$ (C-C) content is clearly observed between the as-received sample, the as-received sample subjected to an argon cluster cleaning (85.6%) and the sample milled with the silicon nitride material (65.8%). In the latter sample, the appearance of the COOH and C-Si components are evident.

Clear differences in the fitting of the C1s peak are observed, showing further graphitization produced with the silicon nitride milling.

Whereas the study of the C1s peak can provide information about the surface graphitization (sp$^2$ content) the analysis of the Fe2p peak will confirm the presence of undesired metal impurities. Iron was detected in the survey spectra of the SC diamond sample milled with the tempered steel. A high resolution scan for the Fe2p peak is shown in Figure 2b. The binding energy of the Fe2p$_{3/2}$ peak was observed at 711.2 eV which corresponds to the core level spectra of Fe$^{3+}$ ions.[44] Although iron was not detected in the SC sample milled with the silicon nitride, a more detailed scan for the Fe2p peak was also performed for this sample for comparison. Figure 2c shows the XPS measurement of this peak in which iron was not detected.

Even though silicon nitride milling is the best method to avoid any metal content in the diamond particles produced, it also has some drawbacks. The silicon nitride material generated in the milling process is difficult to remove. Also, the asymmetry in the C1s peak confirms higher surface graphitization of the particles when compared to the tempered steel milled sample. The sp$^2$ carbon, either in the form of graphitic-like carbon or amorphous carbon, is detrimental for many applications. Attempts to remove the sp$^2$ carbon by conducting an air annealing treatment[45] at 600°C did not result in the complete removal of sp$^2$ carbon.

To confirm the quality of the diamond powders obtained after crushing the SC plates, Raman measurements were performed on the samples.

Figure 4 shows the Raman measurements for the different SC powders. A sharp and clear diamond peak centered at 1332 cm$^{-1}$ can be observed in the samples milled with both tempered steel (figure 4a) and silicon nitride (figure 4b). Furthermore, a small band between 1500 cm$^{-1}$ and 1600 cm$^{-1}$, known as the G-band, can be distinguished in the sample milled with the silicon nitride (figure 4b), confirming the presence of sp$^2$ sites.[46]

The possibility of selecting and controlling the particles' size, highly important for different applications, represents an enormous advantage over commercial nanoparticles. Particles with sizes below 70 nm are desired for quantum applications, with particles as small as 10 nm containing active NV centres.[47] Furthermore, particles with sizes between 50 nm and 100 nm are also suitable for drug delivery bio applications.[48] Particle size distributions can be controlled by combining longer milling times and colloids centrifugation at higher accelerative forces. Figure 5 shows the particle size distributions in the colloids made from tempered steel milled particles as described in the experimental section. Two different characterization methods, DLS and NTA, as well as different accelerative forces to select particle size distributions were used.

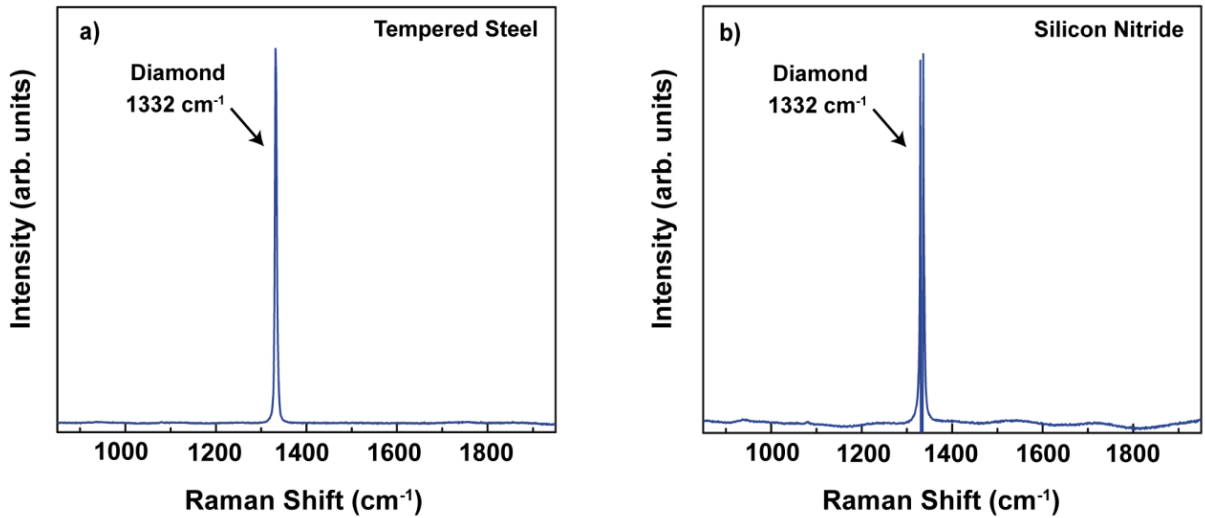

Figure 4. Raman measurements of the powders after the milling, acid cleaning and air annealing processes. a) SC powder after tempered steel milling process and b) SC powder after the silicon nitride milling process.

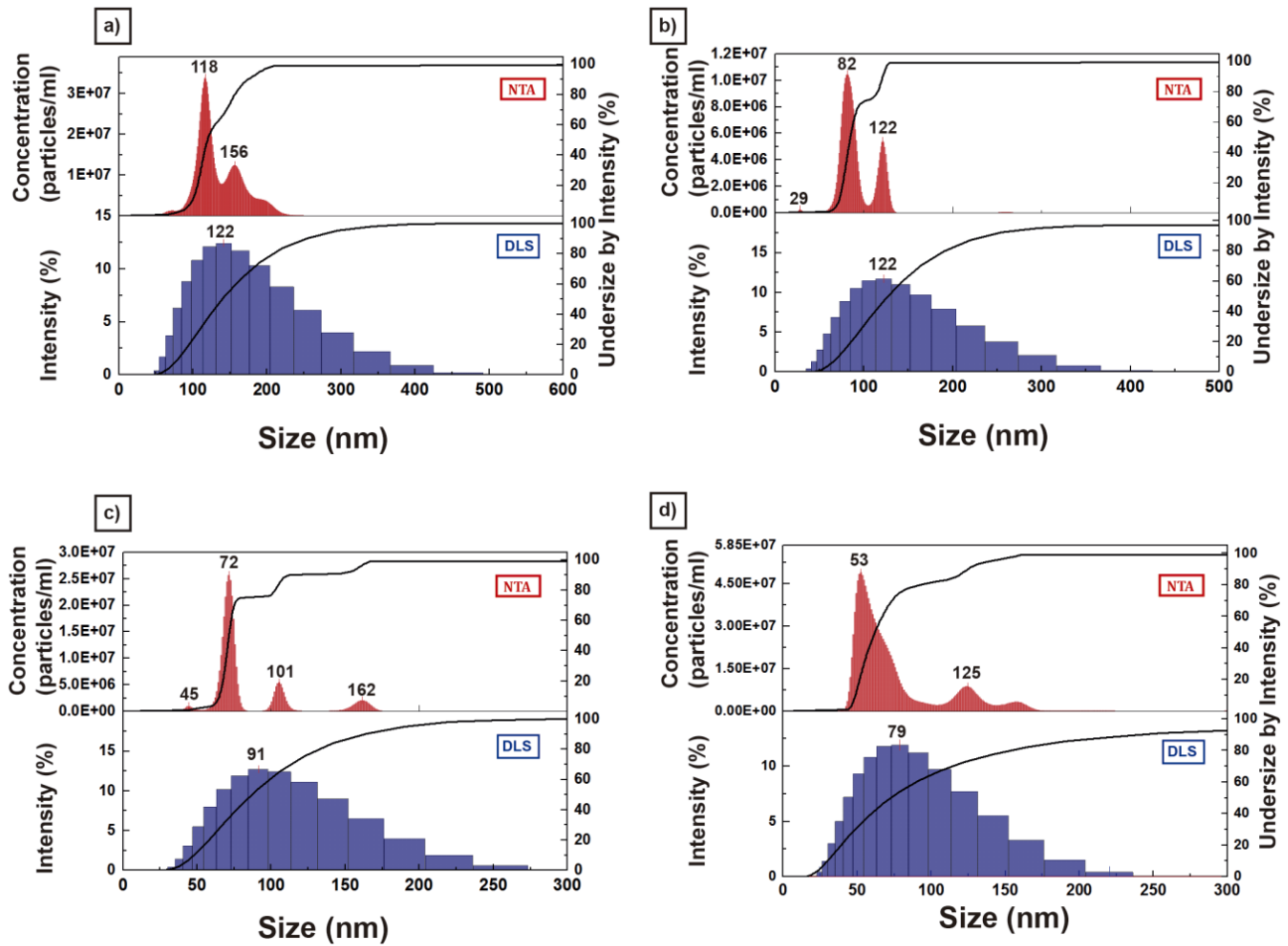

Figure 5. Particles' size distribution of the tempered steel milled powders' solution after centrifugation at different accelerative forces: a) 5000 g b) 10000 g c) 20000 g and d) 30000 g.

In polydispersed colloids, it is important to use various characterization methods. For instance, in DLS, the presence of particles with various sizes can lead to imprecise particle size distributions. This is due to the fact that in DLS, the particle size is determined from intensity fluctuations in the Rayleigh scattering off a volume of the particles. As the intensity of Rayleigh scattering is proportional to $d^6$, where d is the particles diameter, large particles or aggregates can mask the measurement of smaller particles. On the other hand, NTA gives a more precise measure as individual particles can be tracked. Figure 5a shows the distribution of particles after centrifugation at 5000 g. In the NTA analysis graph two different particle size distribution peaks can be differentiated, 118 nm and 156 nm, whereas the particle size distribution is broader for the DLS measurement. This difference increases with increasing centrifuge accelerative force as seen in Figures 5b-d. Three particle size distributions were distinguished at 10000 g (Fig. 5b), 29 nm, 82 nm and 122 nm, but in the DLS measurement a mean value of 122 nm was obtained. Centrifugation at higher rpm, results in smaller fractions of particles with large diameters, and particle size distribution down to 53 nm were recorded after centrifugation at 30000 g.

CONCLUSIONS

In summary, diamond nanoparticles with controlled sizes have been produced following two distinct milling strategies. High quality starting material and the choice of the grinding bowl material will ultimately determine the subsequent potential applications.

Milling with the tempered steel material results in $Fe_2O_3$ presence in the diamond nanoparticles even after the acid cleaning process, which excludes their use in magnetic/spin related applications. Silicon nitride milling is a good choice to ensure metal-free diamond nanoparticles, but results in larger non-diamond contamination, difficult to remove. Although the silicon nitride milled process showed the presence of sodium, the hydroxide cleaning process can be discarded in favor of producing metal free nanoparticles.

ASSOCIATED CONTENT

**Supporting Information**. XPS extended data. "This material is available free of charge via the Internet at http://pubs.acs.org."

AUTHOR INFORMATION

Corresponding Author

* E-mail: GinesL@cardiff.ac.uk and WilliamsO@cardiff.ac.uk

ORCID: 0000-0001-9980-054X

Author Contributions

The manuscript was written through contributions of all authors. / All authors have given approval to the final version of the manuscript.

Notes
The authors declare no competing financial interests.

ACKNOWLEDGMENT

LG, SM and OAW would like to thank the Royal Society International Exchanges Scheme (IE131713) and EU FP7 FET Open "Wavelength tunable Advanced Single Photon Sources". XPS data collection was performed at the EPSRC National Facility for XPS ('HarwellXPS'), operated by Cardiff University and UCL, under contract No. PR16195. GWM is supported by the Royal Society and the EPSRC (EP/J014664/1).